\documentclass[aps,prl,twocolumn,showpacs]{revtex4}

\usepackage{graphicx}

\begin{document}

\title{Coherent ratchets in driven Bose-Einstein condensates}

\author{C.E.~Creffield and F.~Sols}
\affiliation{Dpto de F\'isica de Materiales, Universidad
Complutense de Madrid, E-28040, Madrid, Spain}

\date{\today}


\begin{abstract}
We study the response of a Bose-Einstein condensate to an
unbiased periodic driving potential. By controlling the space and time 
symmetries of the driving we show how a directed current can be 
induced, producing a coherent quantum ratchet. 
Weak driving induces a regular behavior, and both space 
and time symmetries must both be broken to produce a current. 
For strong driving the behavior becomes chaotic
and the resulting effective irreversibility means that it
is unnecessary to explicitly break time symmetry. 
Spatial asymmetry alone is then sufficient to produce the
ratchet effect, even in the absence of interactions, and
although the system remains completely coherent.
\end{abstract}

\maketitle

{\em Introduction -- }
The physics of ratchets, systems that exhibit directed motion
in the absence of an external bias, has undergone extremely rapid 
development in recent years \cite{reimann,hanggi_review}. 
The concept is very general, and ranges from 
new technological
forms of manipulating and directing matter at nanoscale levels,
to understanding how systems in nature such
as biological molecular motors function.
Fundamentally a ratchet must satisfy two essential requirements:
the system must be driven out of equilibrium by an external force, 
and the relevant space/time symmetries, which would otherwise 
forbid the generation of directed currents, must be broken.

A well-known example is provided by a Brownian particle
in a periodic potential. Driving the system from equilibrium
by either pulsing the potential (``flashing ratchet'') or tilting
it (``rocking ratchet'') produces a current if the
spatial or temporal symmetry of the driving force is broken.
In the most commonly studied overdamped regime,
the ratchet current arises from the rectification of random 
fluctuations, and accordingly noise and dissipation are essential 
ingredients. This is not true in general, however,
and surprisingly it has been shown recently 
\cite{ketzmerick,tania,ermann,gong,resonance,interact,poletti} that ratchet 
effects can also occur even in completely 
coherent systems. 

Considerable progress in this direction has been made by considering the
quantum kicked rotor.  Experimentally this system can be realized
extremely well in gases of ultracold atoms held in pulsed
optical lattices. While it was originally thought that ratchet effects 
would only arise in systems with an underlying mixed classical
phase-space \cite{ketzmerick}, recent work has shown that they can 
also arise when the phase-space is globally chaotic.
In Ref. \ \cite{tania} a quantum Hamiltonian ratchet of this type
was studied both theoretically and experimentally,
in which the current arose from the generation
of an asymmetry in the momentum distribution
due to the desymmetrization of the system's Floquet states.
An alternative scheme, developed in the context of
quantum maps \cite{ermann}, is to use interference effects to produce 
an imbalance in the phase-space distribution.
Quantum resonances, where the period of the kicks is matched
to the inverse recoil velocity of the optical lattice, have also
been proposed \cite{gong,resonance} as a means of producing ratchet
accelerators. 

In this work we consider an optically trapped Bose Einstein condensate
(BEC), since the macroscopically protected coherence
and excellent controllability
of these systems make them ideal subjects for investigating
quantum transport effects. Instead of kicking the system \cite{interact,reslen},
we use a smoothly varying potential and so can expect to produce less 
heating effects, which we verify explicitly by evaluating the fraction 
of non-condensed atoms. By choosing a form for the driving which
enables us to separately break space and time symmetries,
we find that we can induce a directed current in a BEC starting from a 
symmetric initial state. This occurs by two distinct mechanisms;
for weak driving the system undergoes regular oscillations,
and both space and time symmetries in the driving must be broken.
Conversely, for strong driving, the system's dynamics
becomes chaotic, and this produces an effective irreversibility
which means it is {\em not} necessary to explicitly break
the time symmetry.

\begin{center}
\begin{figure}
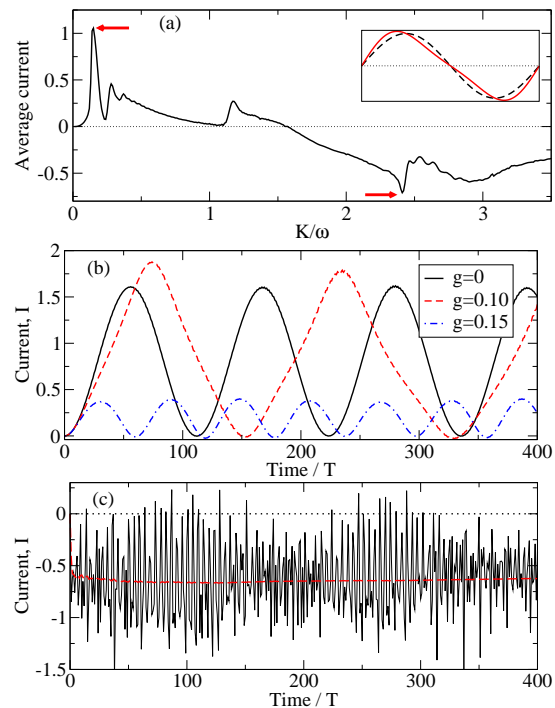

\includegraphics[width=0.4 \textwidth,clip=true]{fig1a.eps}
\includegraphics[width=0.4 \textwidth,clip=true]{fig1b.eps}
\includegraphics[width=0.4 \textwidth,clip=true]{fig1c.eps}
\caption{(a) Time-averaged current (averaged over 100 periods)
a a function of driving strength $K$
for symmetry parameters $\alpha=\beta=0.2$ and nonlinearity $g=0.1$
Clear peaks at $K=0.15$ and $K=2.4$ are marked with
arrows.
Inset: Plot of the driving potential $V(x)$. For
$\alpha = 0$ (black dashed curve) it is symmetric, 
but for $\alpha=0.2$ (solid red curve) it has
an asymmetric sawtooth form.
(b) Current induced for weak driving ($K=0.15$). For
$g=0$ the current exhibits a smooth sinusoidal oscillation; as
$g$ is increased the oscillations are initially enhanced
in amplitude and deviate from sinusoidal form, and 
subsequently become highly suppressed.
(c) Current induced for strong driving ($K=2.4, \ g=0.1$). In contrast
to the above case the current shows rapid quasiperiodic oscillations, which 
nonetheless maintain a stable time-averaged value (red dashed line).}
\label{current} 
\end{figure}
\end{center}

{\em Model -- } 
We consider a BEC confined in a toroidal trap \cite{toroid} with
a cross-section much smaller than the trap's radius, $R$.
The system can thus  be described by an effective
one-dimensional Gross-Pitaevskii equation (GPE)
\begin{equation}
H(t) = - \frac{1}{2} \frac{\partial^2}{\partial x^2}
+ g \left| \psi(x,t) \right|^2 + K \ V(x,t) ,
\label{gpe}
\end{equation}
where $x$ parametrises distance
around the trap, and we measure all energies in units
of $\hbar^2/2 m R^2$.
The short-range interactions between the atoms in the condensate
are described by a mean-field term with strength $g$,
and the condensate is driven by a time-periodic external
potential with zero mean by modulating the amplitude
of the optical potential. The archetypal form of
a ratchet potential \cite{hanggi_review} is 
$V = \sin(x)+\alpha \sin(2x + \phi)$,
where $V$ is symmetric for $\phi=\pi/2$,
and is maximally asymmetric for
$\phi=0, \pi$. We make the unusual choice of
factoring the potential into separate
spatial and time-dependent components of this form 
$V(x,t) = V(x) f(t)$
\begin{eqnarray}
V(x) &=& \sin(x) + \alpha \sin(2 x), \\
f(t) &=& \sin(\omega t) + \beta \sin(2 \omega t).
\label{drive}
\end{eqnarray}
It is important to note that this potential does not correspond to
either a purely rocking or flashing ratchet, and has the appealing
feature of allowing the space and time symmetries of the system to 
be controlled independently. We plot the form of $V(x)$ in the 
inset of Fig. \ref{current}a, and show that for 
non-zero values of $\alpha$ the potential becomes skewed,
breaking both inversion symmetry and shift symmetry 
A static spatial potential of this type has
been recently studied experimentally in \cite{salger}.

{\em Results -- }
To probe the behavior of the system we
evaluate $I(t) = \langle \psi(x,t) | p | \psi(x,t) \rangle$
as a measure of the current flowing in the ring.
As initial condition we choose the zero-current state
$\psi (x,0) = 1/\sqrt{2 \pi}$,
which is convenient for experiment
as it is the ground state of the undriven Hamiltonian,
and so can be prepared by cooling.
We numerically integrate the wavefunction in
time using a split-operator method,
and henceforth we shall take the driving frequency $\omega = 1$.
In Fig. \ref{current}a we show the time-averaged current, 
$\overline{I}$, obtained
by integrating the system over 100 driving periods. For weak driving 
we can note a sharp peak centered at $K=0.15$. As $K$ is
increased from this value the current drops, becomes negative,
and passes through a negative peak at $K=2.4$ This second peak is associated
with an enhancement of (negative) current over a rather broad range
of driving amplitudes.

\begin{center}
\begin{figure}
\includegraphics[width=0.15 \textwidth,clip=true]{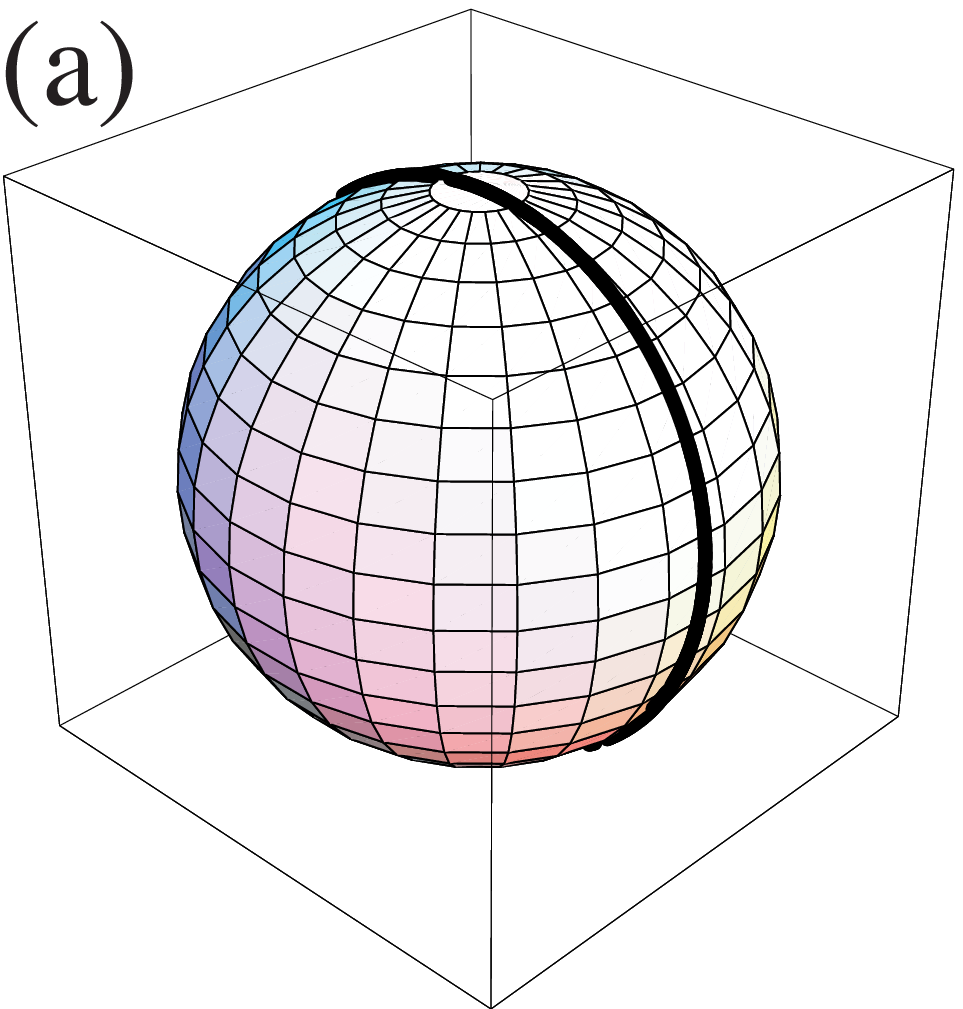}
\includegraphics[width=0.15 \textwidth,clip=true]{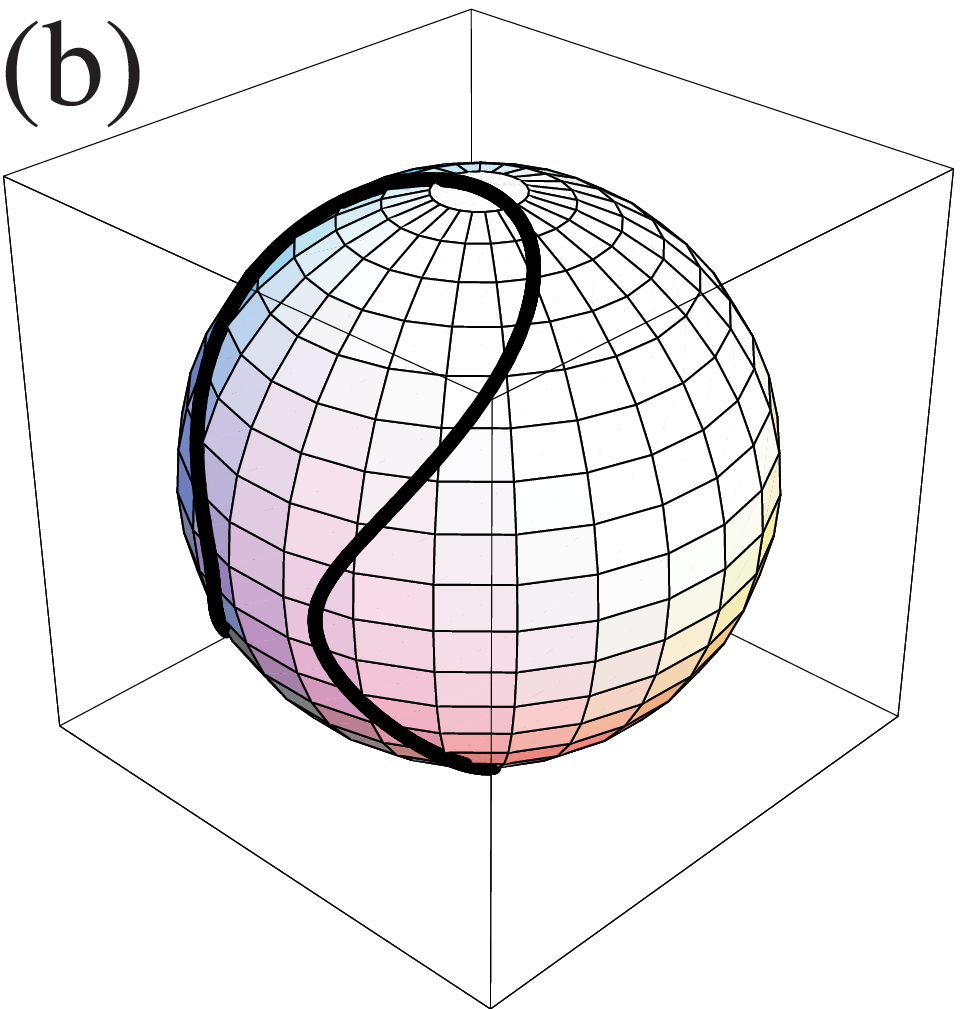}
\includegraphics[width=0.15 \textwidth,clip=true]{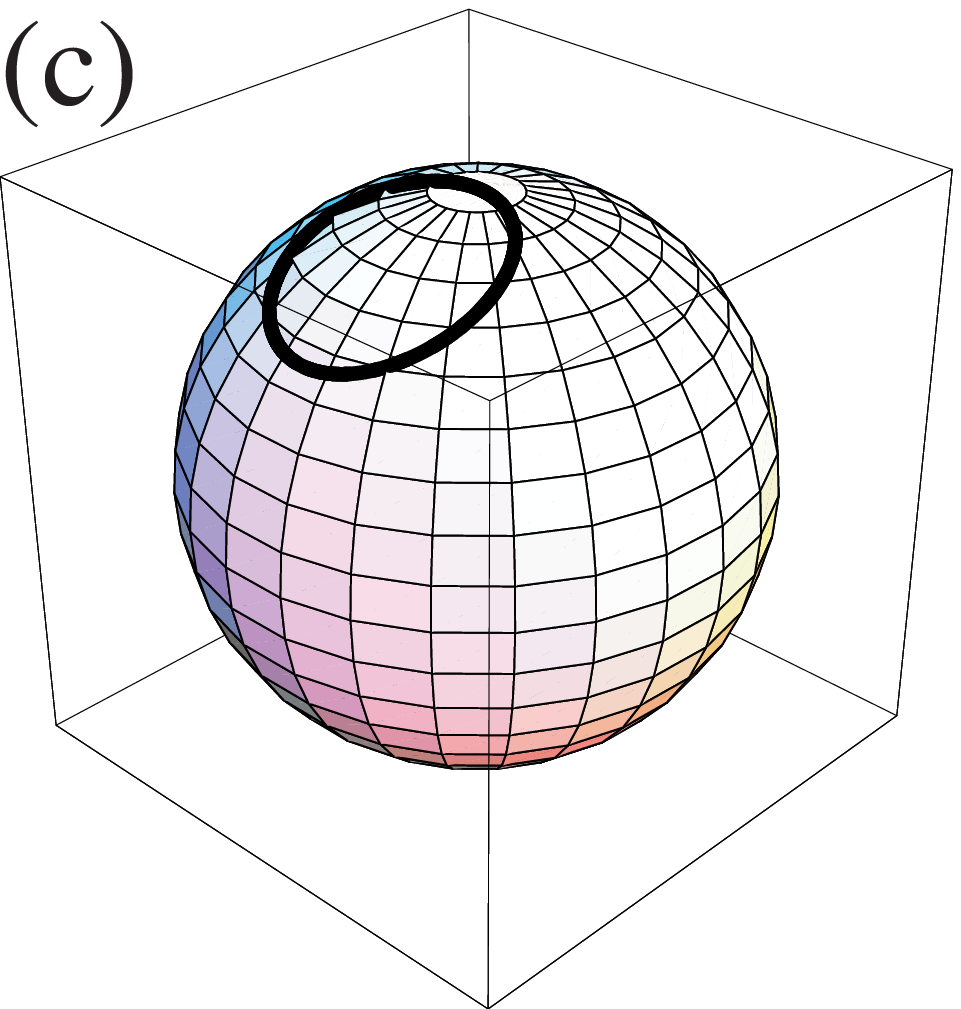}
\includegraphics[width=0.30 \textwidth,clip=true]{fig2d.eps}
\includegraphics[width=0.15 \textwidth,clip=true]{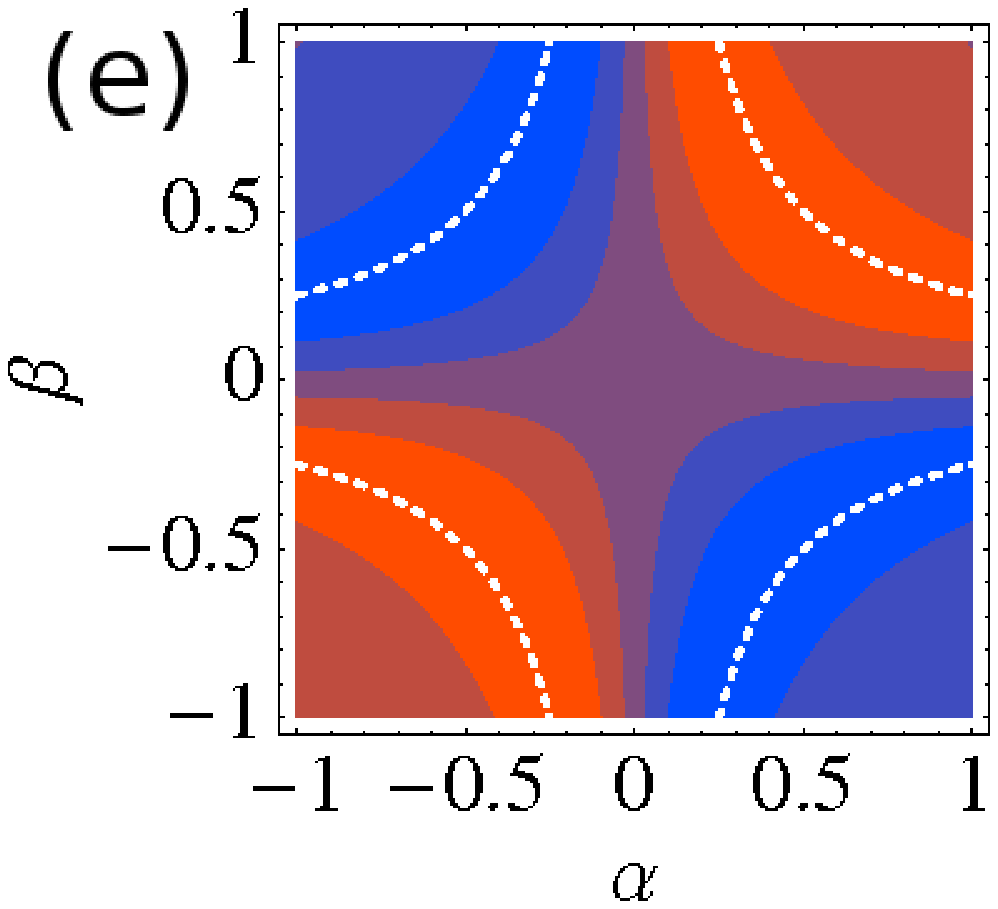}
\caption{Top row: Bloch sphere representation of the the
evolution of the effective two level model (Eq. \ref{2level})
for weak-driving ($K = 0.15$).
(a) $g=0$: the Bloch vector makes a circular orbit in
the $x-z$ plane, corresponding to a sinusoidal
current oscillation.
(b) $g=0.1$: the Bloch vector makes a more
complicated ``figure-of-eight'' orbit,
producing the distortion from the sinusoidal waveform.
(c) $g=0.15$: a phenomenon
analogous to self-trapping occurs, and the Bloch vector is
confined to small orbits in the vicinity of the north pole,
implying an interaction-induced suppression of current. 
(d) The suppression of the current occurs at a very
precisely defined value of $g$. 
The two-level approximation results (dashed red line) compare
well with the simulation of the full GPE (solid black line).
(e) Current produced as a function of the
asymmetry parameters ($g=0.10$). 
The current vanishes for $\alpha$ or $\beta = 0$,
and is positive (negative) in the top-left/bottom right (top-right
/bottom left) quadrants. Contours of constant current
are given by $\alpha \beta = \mbox{cnst.}$ in
agreement with the perturbative result \cite{analytic}; dashed
lines show the current maxima.}
\label{bloch}
\end{figure}
\end{center}

We show the time-dependence of the current for weak driving
in Fig. \ref{current}b. It is clear that the sharp peak in 
$\overline{I}$ is associated with regular 
oscillations of the current. 
For $g=0$ this oscillation is sinusoidal with a period much larger
than $T=2\pi/\omega$. Increasing $g$ initially slightly enhances
the amplitude of the oscillation, while also deforming its waveform.
As $g$ is increased further, however, the
oscillation's amplitude becomes abruptly suppressed (see Fig. \ref{bloch}d).
Examining the evolution of the system in detail
reveals that the oscillation occurs chiefly between the states
$| 0 \rangle$ and $|2 \rangle$, where $| n \rangle$ denotes an
eigenstate of the undriven Hamiltonian with (quantized)
momentum $n \hbar$. While a symmetric driving would equally populate
states with positive and negative momentum, producing
zero current, interference effects cause
the asymmetric form to preferentially drive
the system to $| +2 \rangle$ rather than $| - 2 \rangle$,
thus inducing a net current. A perturbative study of the Floquet
states explains this result \cite{analytic}.

To qualitatively study this phenomenon, let us truncate the wavefunction 
to just the two components of interest,
$\psi(x,t) = A + B \exp(2 i x)$, where
$|A|^2 + |B|^2=1/\sqrt{2 \pi}$. Under the action of the Hamiltonian
(Eq. \ref{gpe}), this yields the equation of motion
for the expansion coefficients $\chi=\left( A,B \right)$
\begin{equation}
i {\dot \chi} =
\left[
- \frac{\alpha K}{2} f(t) \sigma_y
- \left(1 + \frac{g}{2} \left[|A|^2 - |B|^2 \right] \right) \sigma_z \right]
 \chi \ ,
\label{2level}
\end{equation}
where $\sigma_j$ are the Pauli
matrices. We may now visualize the time-evolution of the system
using the Bloch sphere representation, where the
north / south poles correspond to occupation of the 
states $| 0 \rangle \ / \ | 2 \rangle$.
For $g=0$ the Bloch vector will evolve under the influence
of a fictitious magnetic field $B_y = \alpha K f(t)$,
and so will simply make a Larmor orbit in the $x-z$ plane as
shown in Fig. \ref{bloch}a. This corresponds to the sinusoidal
oscillation displayed in Fig. \ref{current}a. For a larger
value of $g$ the Bloch vector will execute a more complicated 
``figure-of-eight'' motion under the combined influence
of $B_x$ and $B_z$, shown in Fig. \ref{bloch}b,
producing the non-sinusoidal current oscillations seen in Fig. \ref{current}b.
When $g$ is increased further the magnitude of the
fictitious field component $B_z$ is enhanced, until the Bloch
vector is confined to making rapid circular orbits (Fig. \ref{bloch}c) 
near the north pole, in a process directly analogous to the 
non-linear phenomenon termed self-trapping \cite{oberthaler}.
These small orbits correspond to the low amplitude,
high frequency oscillations seen in Fig. \ref{current}b for
large values of $g$. In Fig. \ref{bloch}d we show in detail 
how this transition occurs by plotting the
amplitude of the current oscillations as $g$ is increased, 
and demonstrate that this simple model indeed captures
the main features of this effect.

\begin{center}
\begin{figure}
\includegraphics[width=0.4\textwidth, clip=true]{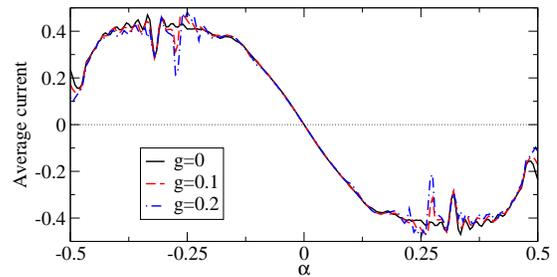}
\caption{Ratchet current produced for strong driving
($K=2.4$) for a system with
{\em no} temporal asymmetry (i.e. $\beta=0$).
Non-zero spatial asymmetry ($\alpha \neq 0$)
is required to induce a current, and the direction of the
current depends on the sign of $\alpha$.
The magnitude of the current depends only
weakly on the nonlinearity $g$.}
\label{symm}
\end{figure}
\end{center}

We now consider the case of strong driving ($K=2.4$).
In this regime the initial state is driven to a larger
number of excited levels, and consequently the two level approximation
is no longer valid. The generated current, shown in Fig. \ref{current}d, 
instead shows an irregular, quasiperiodic character
corresponding to the large number of frequency components present.
Despite its jagged appearance the current converges to a 
stable time-average after just a few tens of driving periods.
The ratchet current thus arises from a completely
different mechanism to before; instead of a regular oscillation
between two momentum states, the system now evolves to a
superposition of many momentum eigenstates, which crucially
has an asymmetric momentum distribution \cite{tania}.
This produces a dramatic difference
between the symmetry dependence of the current in the weak and
strong driving regimes. The symmetry properties of this form of 
driving were analyzed in Ref. \ \cite{denisov} for a non-interacting
system, and it was concluded that 
{\em both} space and time symmetries needed to be broken
to produce a ratchet current. This is indeed the case for weak driving,
and in Fig. \ref{bloch}e we can clearly see
that the current vanishes for $\alpha, \beta = 0$. 
In contrast, when the system is driven strongly
only $\alpha \neq 0$ is required, that is, it is sufficient to
{\em just} break the spatial symmetry,
an effect also noted in quantum resonant ratchets \cite{spatial}. 
In Fig. \ref{symm} we show the current produced for strong driving
when the temporal asymmetry $\beta$ is set to zero.
Away from $\alpha=0$ we obtain a clear ratchet current
whose direction depends on the sign of $\alpha$. 
This occurs in analogy to the production of
a ratchet current in a non-Hamiltonian system \cite{ferruccio}, 
but instead of dissipation it is the quasiperiodic evolution
of the system which produces the effective
irreversibility in time \cite{mieck}.
This contrasts sharply with the results obtained in
Ref. \ \cite{poletti}, where instead the interaction was argued
to play the role
of breaking the time-reversal symmetry. As a result, the ratchet
current we obtain has only very weak dependence on $g$ (as shown in 
Fig. \ \ref{symm}), and occurs even in the non-interacting case.

As well as driving the dynamics of the condensed atoms, the
potential also has the effect of exciting atoms out
of the BEC to form a thermal cloud. This depletion eventually
leads to the destruction of the BEC, and seeing
how rapidly this occurs allows us to assess the stability of the BEC
under driving \cite{reslen}. 
This can be done by making the Bogoliubov approximation,
and linearizing the GPE about its ground-state. Following
the Castin-Dum formalism \cite{castin}, the mean number of
non-condensed atoms at zero temperature is given by
$N(t) = \sum_{k \neq 0} \langle v_k(t) | v_k(t) \rangle$,
where $(u_k, v_k)$ are the amplitudes
of the Bogoliubov quasiparticle operators with quantized momentum $k$.
This approximation is valid provided that the number
of particles excited from the condensate is small
compared to the condensate itself.

\begin{center}
\begin{figure}
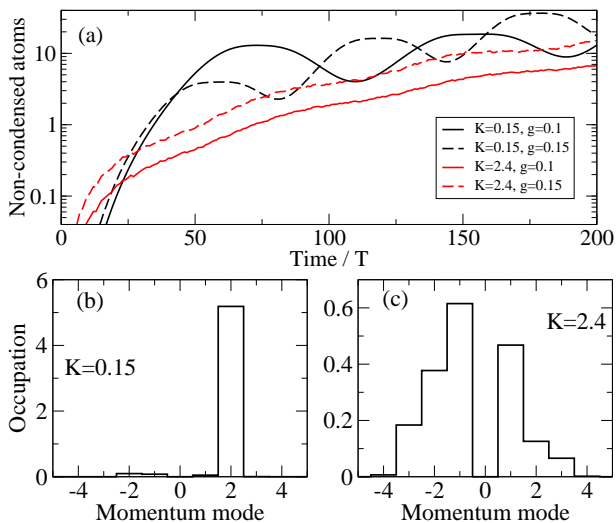

\includegraphics[width=0.45\textwidth, clip=true]{fig4a.eps}
\includegraphics[width=0.45\textwidth, clip=true]{fig4b.eps}
\caption{(a) For weak driving ($K=0.15$, upper black
curves) the number of non-condensed atoms exhibits
significant oscillations, arising from the oscillation
of atoms in the condensate. For strong driving
($K=2.4$, lower red curves) the number rises monotonically
with a similar approximately power-law rate of increase.
(b) Occupation of non-condensed modes after 100 driving
periods for weak driving ($g=0.1$). This distribution mimics the
spectrum of the condensate, with a large peak
at $n=+2$.
(c) As for (b) but for strong driving. The distribution
of momentum is again asymmetric, but many modes are populated.}
\label{cloud}
\end{figure}
\end{center}

In Fig. \ref{cloud}a we show the time development of the non-condensed atoms
for both weak and strong driving. In all cases the
number of atoms does not increase exponentially,
but instead follows an
approximate power-law, $N \propto t^\gamma$ with $\gamma \simeq 1.9$,
indicating that under the driving the condensate does not
exhibit dynamical instability. It is interesting to note
that for weak driving the production of non-condensed atoms
exhibits marked oscillations. These are a consequence
of the large, slow current oscillations that the driving 
generates; the regular sloshing motion periodically produces
peaks in the density distribution of the condensate, causing
an enhanced emission of particles at those times and locations.
As the density distribution can be directly imaged in experiment,
this provides a convenient means to study the dynamics
of the BEC in this regime.
The momentum distribution of the non-condensed atoms also mimics
the behavior of the condensate. In Fig. \ref{cloud}b we
see that weak driving essentially only excites a
single Bogoliubov mode, with momentum $k=+2$.
As the driving is not resonant with the Bogoliubov frequency,
however, this mode does not grow exponentially with time \cite{reslen}
and so dynamical instability is avoided.
Conversely, for strong driving a
larger number of Bogoliubov modes are excited, and
the momentum distribution is 
strongly asymmetric, again resembling the momentum spectrum
of the condensate.

{\em Conclusions -- }
We have investigated the dynamics of a BEC under
a periodic driving potential. In the weak driving regime
a ratchet current can be generated by breaking both space and time
symmetries, and inducing slow, regular oscillations of
the condensate. For
strong driving a more complicated quasiperiodic dynamics is
induced, in which chaos acts to eliminate
the long term memory of the system, thus mimicking thermal noise.
Accordingly only the spatial symmetry then needs to be broken 
to produce a ratchet current. 
This current should remain stable over
timescales comparable to the system's Ehrenfest
time, which scales logarithmically with the number
of atoms in the condensate \cite{shep},
as also suggested by numerical results presented in Ref. \ \cite{poletti}. 
For a condensate of $10^5$ atoms
we estimate this time to be of the order of 50 driving periods.
These results expose a new vista of possibilities in
manipulating the interplay between the driving potential,
interactions and symmetry breaking to induce directed
transport in quantum coherent systems.

\bigskip

This work was supported by MICINN (Spain) through grant
FIS-2007-65723 and the Ram\'on y Cajal Program (CEC).
The authors thank Ferruccio Renzoni, Tania Monteiro, and
Peter Reimann for stimulating discussions.

\end{document}